\documentclass[10pt,final,journal,twoside]{IEEEtran}
\usepackage[T1]{fontenc}
\usepackage[mathscr]{eucal}
\usepackage{ifpdf}
\usepackage{cite}
\usepackage{graphicx}
\usepackage[cmex10]{amsmath}
\usepackage{amssymb}
\usepackage{algorithmic}
\usepackage{units}
\usepackage{setspace}
\usepackage{algorithm}
\usepackage{array}
\usepackage[tight,footnotesize]{subfigure}
\usepackage{amsthm}
\usepackage{multirow}
\usepackage{enumerate}
\usepackage{color}
\usepackage{makecell}

\newcommand*{\affmark}[1][*]{\textsuperscript{*}}
\begin{document}
\title{Mobile Edge Computation Offloading Using Game Theory and Reinforcement Learning}
\author{
\IEEEauthorblockN{Shermila Ranadheera, Setareh Maghsudi, and Ekram Hossain\\}
\thanks{S. Ranadheera and S. Maghsudi contributed equally to this work. S. Ranadheera and E. Hossain are with the Department of Electrical and Computer Engineering, University of Manitoba, Winnipeg, Canada (e-mail: ranadhes@myumanitoba.ca, Ekram.Hossain@umanitoba.ca). S. Maghsudi is with the Department of Electrical Engineering and Computer Science, Technical University of Berlin, Berlin, Germany (e-mail:maghsudi@tu-berlin.de). 
}
}
\maketitle
\begin{abstract}
Due to the ever-increasing popularity of resource-hungry and delay-constrained mobile applications, the computation and storage capabilities of remote cloud has partially migrated towards the mobile edge, giving rise to the concept known as Mobile Edge Computing (MEC). While MEC servers enjoy the close proximity to the end-users to provide services at reduced latency and lower energy costs, they suffer from limitations in computational and radio resources, which calls for fair efficient resource management in the MEC servers. The problem is however challenging due to the ultra-high density, distributed nature, and intrinsic randomness of next generation wireless networks. In this article, we focus on the application of game theory and reinforcement learning for efficient distributed resource management in MEC, in particular, for computation offloading. We briefly review the cutting-edge research and discuss future challenges. Furthermore, we develop a game-theoretical model for energy-efficient distributed edge server activation and study several learning techniques. Numerical results are provided to illustrate the performance of these distributed learning techniques. Also, open research issues in the context of  resource management in MEC servers are discussed.
\end{abstract}
\section*{Introduction}
\label{Sec:Intro}
\begin{figure}[!htb]
\centering
\includegraphics[width=0.45 \textwidth]{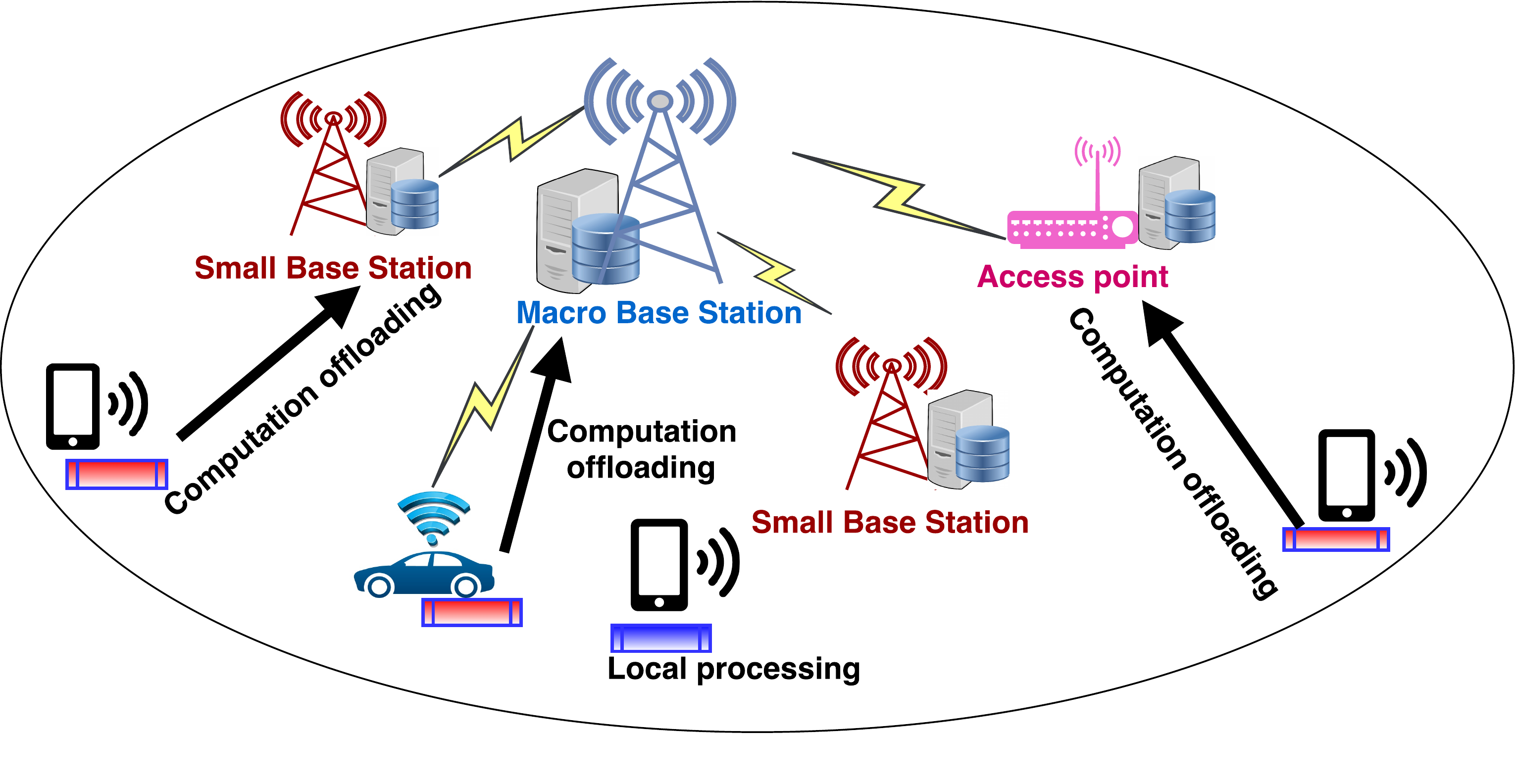}
\caption{Mobile edge computing and computation offloading.}
\label{Fig:MEC}
\end{figure}

Due to restricted battery power, memory and computational capacity, mobile devices face challenges in executing delay-sensitive and resource-hungry mobile applications such as augmented reality and online gaming. Mobile Edge Computing (MEC) is foreseen as a remedy to alleviate this problem. In MEC, the mobile edge is enhanced with analysis and storage capabilities, possibly by a dense deployment of computational servers or by strengthening the already-deployed edge entities such as small cell base stations. Consequently, mobile devices are able to offload their computationally expensive tasks to the edge servers while requesting some specific quality of service. This process, referred to as computation offloading, is feasible due to the fact that edge servers are deployed in close proximity of mobile users, specifically in comparison to the remote cloud servers. An illustration of MEC is provided in Fig. \ref{Fig:MEC}.

Notwithstanding its numerous perks, MEC suffers from some short-comings that should be addressed for the concept to become realizable. Most importantly, the limited radio and computational resources of mobile edge servers shall need to be efficiently utilized so that the users' quality of service requirements are met with the minimal effort. Moreover, from the mobile devices' perspective, the consumed energy shall be minimized. The problem becomes aggravated when the randomness and dynamics of wireless networks are taken into account. The factors that contribute to this issue include, but are not limited to, users' mobility, random channel quality, time-varying and random task-arrival, non-deterministic energy resources (for instance in case of energy harvesting), and other similar variables.   

Naturally, the aforementioned challenges cannot be addressed by conventional centralized resource allocation schemes, since such mechanisms necessitate the availability of global information at a central node. This is infeasible to acquire in ultra-dense distributed networks. Moreover, the computational complexity becomes overwhelming as well. As a result, it is vital to develop distributed and autonomous approaches, where the individual mobile devices and mobile edge servers make decisions for the system to settle at efficient and stable operating points. Therefore, game theory and reinforcement learning are considered as two mathematical tools with great potential to address such problems. 

%
Game theory is well-established as a classic tool to mathematically model the wireless resource allocation problems. Based on the fact that many of the wireless resource allocation problems can be reduced to distributed decision making problems, game theory becomes an ideal fit. Game theory focuses on strategic interactions among players and thus eliminates the need for a central controller which is a major advantage. As it is well-known, game theory has two main branches: non-cooperative and cooperative. Non-cooperative game theory studies the interactions of rational and self-interested players that compete against each other, and the goal is to achieve an efficient equilibrium point. Important notions of equilibrium include Nash, correlated and Walrasian equilibrium. Cooperative game theory promotes a cooperative behavior that is supposedly beneficial to all agents; a well-known example is coalitional games. In such games, players form coalitions, and enforce cooperative behavior in each coalition, so as to maximize the value of the coalition, which is regarded as a utility measure. In addition to coalition formation, there are also some other types of games based on strategic cooperation. In summary, game theory offers a variety of game models in which, each game has its own distinct set of properties, that make them suitable for different types of decision making problems based on the context.

Minority game (MG) is a type of non-cooperative game that can be used to model distributed resource allocation problems in MEC. More precisely, MG is a congestion game in which, an odd number of selfish players choose between two actions in the hope of maximizing their own payoffs. Only the players landing in the minority are rewarded. The players neither communicate with each other nor they have any information about the actions of the other players. Therefore, decision making is almost entirely autonomous. The winning action is broadcast to all players at the end of each round of play which is the only external information provided. This lack of availability of information calls for adaptive methods to be used in order to determine the best action to be chosen in the next round. To this end, the existing literature offers a significant number of learning methods that include reinforcement learning methods and stochastic strategies. These methods help players improve the coordination among their actions and achieve better social and individual welfare, by forming larger minorities \cite{Catteeuw2012}. 

The applicability of MG as a tool for modeling resource allocation problems is quite obvious. In most congestion problems, being in the minority tends to be more beneficial. Many wireless resource allocation problems are in fact congestion problems, where a number of users compete for a limited resource. If the resource is uncrowded, users are rewarded, which is analogous to the minority being rewarded in an MG. For instance, consider the previously mentioned MEC system where, the offloading users attempt to utilize the limited amount of available computational resources of the edge servers. This scenario essentially maps to a congestion problem, since the utility of each user depends on the number of users using the same resource. Hence, such problems can be easily modeled using MG. The advantages of MG include simple implementation, low overhead, and scalability to large set of players, which are of vital importance in a dense wireless MEC system. More details on MG can be found in \cite{7891693}. Later in this article, we provide an MG model for distributed server activation problem in MEC where we compare the performance of several learning algorithms used in MG.

Reinforcement learning is another well-known technique applicable to distributed decision making. In reinforcement learning, autonomous agents learn the best action by using the rewards and penalties received in each round of play. Since agents do not know which action is the best, they learn by balancing exploration of unknown actions and exploitation of the current knowledge of used actions. In other words, agents use trial and error approach to maximize their utilities over the horizon. Some well-known reinforcement learning methods include Q-learning, learning automata and Roth-Erev learning. Reinforcement learning mechanisms are very-well suited for learning in MG, since adaptation to the collective action of the other agents in the presence of information scarcity can be achieved using such methods \cite{Catteeuw2012}. 

In this article, first we provide a concise review of the state-of-the-art in computation offloading and efficient resource allocation of edge servers, with an emphasis on the solutions developed by using game theory and reinforcement learning. 
Then, we explore the research outlook and open problems. 
To this end, we formulate an example distributed server activation problem as a minority game and apply reinforcement learning to solve the game. We present numerical results on the performance of the different learning techniques.  
\section*{State-of-the-Art}
\label{Sec:State}
In this section we briefly explore the cutting edge research in the area of computation offloading and resource management for MEC. In doing so, we focus on computation offloading and resource management methods that are developed based on game theory 
and/or reinforcement learning. Note that a comprehensive survey of the state-of-the-art is out of the scope of this article, and our goal is to capture the research trend by reviewing some exemplary research works. 


\subsection*{Computation Offloading}
\label{subsec:Offloading}
In \cite{7500395}, the authors consider a multi-cell, quasi-static environment, and cast the computation offloading problem as a dynamic sequential game. They further establish the existence of Nash equilibrium and develop a distributed convergent offloading scheme. In \cite{7636777}, the authors consider the offloading problem with the set of mobile devices varying randomly during the offloading period. The problem is modeled using a stochastic game framework, which is afterward shown to be equivalent to a potential game. The existence of Nash equilibrium is proved and a stochastic learning algorithm is developed. For cloud-enhanced vehicular networks with edge computing capability, an offloading mechanism based on a Stackelberg game is proposed in \cite{7997360}. The servers and the offloading vehicles are modeled as the leaders and the followers, respectively. Similar to the aforementioned references, the existence of Nash equilibrium is proved and a distributed algorithm is designed that maximizes the edge server's utility while meeting the tasks' latency constraints. In \cite{7028516}, the authors investigate the multi-user offloading decision making problem in a dynamic environment, where users' states and offloading requests are time-variant. The number of tasks offloaded to each server (machine) is modeled as an a priori unknown time-varying Markov process. The authors then formulate the offloading problem as a Markov decision process. Online learning algorithms are developed to solve for the optimal offloading policy for both centralized and decentralized scenarios. 
\subsection*{Radio and Computational Resource Management}
\label{subsec:Resource}
In \cite{8009942}, the authors use coalitional game theory to solve a resource allocation problem in MEC-enabled IoT networks with software-defined network (SDN) capability. In such a network, delay sensitive tasks are offloaded to the edge servers by the IoT applications. The developed game-theoretical framework is guaranteed to adaptively provision the available computational resources in the MEC servers in order to satisfy the quality of service requirements of IoT applications. Moreover, a deterministic algorithm is proposed to minimize the task processing cost and the latency. Reference \cite{7973020} investigates joint offloading decision making and dynamic edge server provisioning in an offloading mobile edge network with energy harvesting capability. They model the problem as a Markov decision process. A reinforcement learning algorithm is developed for offloading computation jobs and activating edge servers while minimizing the overall cost and delay. The authors of \cite{8029256} propose a resource allocation mechanism using auction theory. Therein, service providers in the mobile edge network design contracts with the edge node infrastructure providers. The contracts enable the edge servers to efficiently provision their assigned computational resources and to schedule the offloaded tasks in a way that the latency is minimized. In \cite{8061008}, the focus is on a dynamically-changing vehicular networks with MEC capabilities including computation and caching. A network operator allocates computation, caching and network resources to the vehicles for different vehicular applications. To address high complexity, the authors develop a deep reinforcement learning algorithm based on deep Q-learning. 
\begin{table*}[t]
\centering
\caption{A comparison of the state-of-the-art}
\begin{tabular}{|c|c|c|c|} 
\hline 
\textbf{Reference} & \textbf{Objective} & \textbf{Model} & \textbf{MEC type} \\ \hline
\cite{7500395} & minimize users' energy and latency cost & dynamic sequential game& quasi-static \\ \hline 
\cite{7636777} & minimize users' energy and latency cost & stochastic game & dynamic \\ \hline
\cite{7997360} & maximize utilities of users and servers & Stackelberg game & vehicular  \\ \hline 
\cite{7028516} & minimize unprocessed offloading requests & Markov decision process & dynamic  \\ \hline
\cite{8009942} & optimize resource usage and QoS guarantee & coalitional game & edge IoT  \\ \hline
\cite{7973020} & minimize overall cost and latency & Markov decision process & energy harvesting MEC \\ \hline
\cite{8029256} & minimize latency & auction theory & dynamic workload arrival \\ \hline
\cite{8061008} & efficient resource allocation & deep Q-learning & vehicular \\ \hline
\cite{Shermila17:Corr} & minimize servers' energy and QoS guarantee & minority game & random   \\ \hline
\end{tabular}
\label{tab:1}
\end{table*}
\section*{Research Outlook}
\label{Sec:Outlook}
Despite its great potential in improving the latency and energy consumption, realizing the concept of MEC is associated with a variety of challenges. In particular, decision making for computation offloading as well as joint radio-computational resource allocation are challenging. The challenge mainly arises due to resource scarcity and distributed nature of MEC, as well as the uncertainty and randomness in wireless networks. This includes, but is not limited to, the randomness in channel quality and the amount/type of offloaded tasks. In what follows, we briefly discuss some important problems, including computation offloading and few other closely-related issues. We also investigate the ability of game theory and reinforcement learning to address the challenges and obtain efficient solutions.

\textbf{Computational Resource Allocation:} As a result of being deployed at the edge, MEC suffers from restrictions of computational resources, in particular when compared to the central mobile cloud computing. As a result, it becomes imperative to allocate the limited resources in an efficient manner. This includes, but is not limited to, MEC server activation and scheduling, load balancing, request management, task allocation, and the like. Such problems can be in particular addressed by cooperative games, where a set of entities form coalitions to achieve a specific goal, and then share the reward. Moreover, by combining reinforcement learning with game theory, the uncertainty and lack of prior information can be addressed.

\textbf{Radio Resource Allocation:} Enhancing the wireless network with MEC complicates the radio resource management. For instance, the necessary uploading and downlinking of task-related data results in radio bandwidth consumption and interference. Consequently, smart bandwidth allocation shall need to be performed for mobile devices/servers. Moreover, the energy consumption at the servers should be kept at the minimum. To increase energy efficiency, servers might share the energy resources and/or harvest ambient energy. Such remedies however introduce uncertainty in the system, in contrast to using deterministic power resources such as a grid. The problem can be addressed by using models from cooperative game theory and reinforcement learning.

\textbf{Computation Offloading:} While the allocation of computational resources is performed on the MEC servers' side, mobile devices decide about computation offloading. In essence, each device decides which and what part of every task shall be offloaded to an edge server. In some cases, the specific server to which the task is uploaded can be determined by the device as well. Moreover, mobile devices might be able to demand a specific quality of service guarantee. Naturally, mobile devices might compete with each other for limited computational services, whereas each server would compete with others to increase its number of offloaded tasks. Moreover, a conflict arises between the set of servers and the set of devices, since the latter requests low prices for services, whereas the former benefits from high service prices. All such scenarios can be modeled and solved by using competitive games and models from economic markets. As before, a convergence to an efficient solution can be achieved by performing the game repeatedly and learn from the outcomes.

\textbf{Information-Centric MEC:} Inspired by the concept of caching of popular files, in information-centric MEC, the data 
and/or services can be saved at different edge servers to promote an efficient computation. In fact, by using this concept, the amount of data which should be uploaded/downlinked dramatically reduces. Naturally, not all the data/services can be cached at every server. In addition, the service demand for users might change over time. Thus the problem to address is as follows: How much and which data/services shall be saved at each server? In addition, the servers should be motivated to cooperate with each other, so that if necessary, the tasks/data/services can be exchange among servers. Such problems can be addressed by using cooperative game theory, repeated auctions, and exchange economy.

\textbf{Economics of MEC Server Virtualization:}  Mobile network operators (MNOs) or service providers (SPs) may lease the MEC servers/resources from  infrastructure providers (InPs). The InPs then will need to virtualize their MEC resources among different MNOs/SPs. The economics of the virtualization of MEC resources can be modeled and analyzed using game theory models. As an example, for a scenario with multiple InPs and multiple MNOs/SPs, a multi-leader and multi-follower Stackelberg game model can be formulated to determine the equilibrium prices that the MNOs/SPs need to pay to the InPs. In a more general scenario, virtualization of MEC resources/servers can be combined with virtualization of other resources including infrastructures (e.g., base stations), spectrum resources, as well as caching storage. Modeling and analysis of such a general virtualized network under users' quality of experience (QoE) constraints is an interesting research problem. 

\section*{Energy-Efficient Computation Offloading}
\label{system_model}
Following the previous discussions, in this section we formulate an energy-efficient server activation problem. We then solve the formulated problem by using minority games in conjunction with reinforcement learning. 

Consider a virtual pool of $M$ edge computational servers, gathered in a set $\mathcal{M}$. At consecutive rounds $t=1,2,...$, the pool receives a fixed number of offloaded computational tasks to perform. Tasks are delay-sensitive with some execution deadline. At every time slot $t$, $c_{t}$ servers are active and the offloaded computing tasks are equally divided among the active servers. On one hand, since each task requires a random time to be performed, the number of servers should be large enough to guarantee an acceptable user experience. On the other hand, initial activation of a server, as well as performing each task, require some fixed amount of energy. Every active server is reimbursed for its performed tasks. Thus, the number of tasks per servers shall be large enough to insure an acceptable revenue. Based on this trade-off, one can determine the required number of active servers at each offloading round so that (i) the system is energy-efficient; and (ii) the user's quality of experience is satisfactory with high probability. We show this threshold number with $c_{\textup{th}}$, and take it as given in this paper. An example calculation of $c_{\textup{th}}$ can be found in \cite{Shermila17:Corr}.

In a distributed MEC system, prior to task arrival, every server independently decides whether to 
\begin{itemize} 
\item accept computation jobs (\textit{active} mode); or 
\item not to accept any computation job (\textit{inactive} mode). 
\end{itemize} 
That is, each server has two possible actions. Based on the discussion above, desired is to have $c_{t}=c_{\textup{th}}$ at every offloading round $t=1,2,...$. In what follows, we model this problem as a minority game and use a variety of learning algorithms to solve the game. 
\subsection*{Modeling the Problem as a Minority Game}
\label{MGmodel}
A MG can model the interaction among a large number of players competing for limited shared resources. In a basic MG, the players select between two alternatives and the players belonging to the {\em minority} group win. The minority is typically defined using some cut-off value. The collective sum of the selected actions by all players is referred to as the \textit{attendance}. 

We model the formulated server mode selection problem as an MG, where the $M$ servers represent the players, with a cut-off value 
$c_{\textup{th}}$ for the number of active servers. The game is repeated at consecutive rounds. The action of an agent $i$ at time $t$ is denoted by $a_{i,t} \in \{0,1\}$. A server being active and inactive correspond to $a_{i,t}=1$ and $a_{i,t}=0$, respectively. Thus $c_{t}$ is equivalent to the attendance. If $c_{t} \leq c_{\textup{th}}$, active servers are winners, and each receives a unit reward. In contrast, 
$c_{t}>c_{\textup{th}}$ promotes inactive servers as winners, yielding a unit reward for each of them. We use $\sigma$ to denote the standard deviation of the attendance value $c(t)$. We define the \textit{volatility} as $\sigma^2/M$. Note that volatility corresponds to the inverse global efficiency (social welfare) of the MG, since smaller volatility implies larger minority size, thereby larger number of satisfied agents. It should be mentioned that zero volatility is considered as the Nash equilibrium of MG.
\subsection*{Distributed Learning Algorithms}
\label{learningalgorithm}
In an MG, the agents apply an algorithm to learn the best action to be played in the next round of play. In the seminal studies of MG, a distributed learning algorithm is introduced, where each agent plays MG with the help of a given set $\mathcal{S}$ of strategies. Each strategy $s \in \mathcal{S}$ specifies an action to be played for every possible history data string. The agents evaluate their strategies by scoring them for the accuracy of their predictions as the game evolves, and use the strategy with the highest score in each round \cite{Challet:2014}. Apart from this seminal mechanism, a variety of learning algorithms are available in the MG literature that can be used by agents to learn the best action. Many of these algorithms fall into the category of reinforcement learning, where the learners balance the exploration-exploitation trade-off in order to maximize their utilities. In addition, learning methods based on stochastic strategies are also available where agents choose their actions with some probability. In what follows, we introduce some of these algorithms and their applicability in an MG setting. 
\begin{itemize}
\item \textbf{Exponential Learning:} In \cite{Marsili2000522}, exponential learning is applied in MG. Each agent is given $S$ strategies, and the agent scores each of these strategies based on the accuracy of its prediction of the winning action. Each agent $i$ selects a strategy $s$ with some probability $p_{i,s}(t)$, defined as:
%
$p_{i,s}(t)=e^{\gamma_i V_{s,i}(t)}/(\sum_{s'=1}^{S}e^{\gamma_i V_{s',i}(t)})$, 
%
where $V_{s,i}(t)$ is the score of strategy $s$ at time slot $t$. Moreover, $\gamma_i$ is the \textit{learning rate} of each agent. Note that, $\gamma_i=\infty$ corresponds to selecting the strategy with the highest score which is the seminal MG learning algorithm.
\item \textbf{Q-Learning:} In \cite{Catteeuw2012}, Q-learning is applied in an MG, where each agent keeps track of the Q-value of two actions. Every agent $i$ uses the following rule to update the Q-values, where $U_{i,a}$ is the utility received by agent $i$ as a result of some action $a$. This rule makes use of the utility information ($U_i$) possessed by the agents in order to learn the best action (i.e., exploitation of the available information). The Q-learning in MG is two fold; (i) Q-values are determined for the two actions (we refer to this as \textit{Action-based Q-learning }) and (ii) Q-values are determined for agents' strategies (we refer to this as \textit{Strategy-based Q-learning}). In the second scenario, an agent keeps track of the Q-values for each of her \textit{strategies}:
\begin{align*} 
\label{Qvalue}
Q_{i,a}(t+1)=\begin{cases}
Q_{i,a}(t)+\gamma_i(U_i(t)-Q_{i,a}(t)),\ a_{i,t}=a \\
Q_{i,a}(t),\ \text{O.w.} 
\end{cases}
\end{align*}
Given Q-values and some $\epsilon>0$, every agent selects the action with the highest Q-value with probability $1-\epsilon$, and with probability $\epsilon$ selects an action uniformly randomly (i.e., exploration). 
\item \textbf{Adaptive Strategy:} Authors in \cite{Lam:2007:ASM} developed an adaptive learning strategy for MG. Therein, for each actions $a$, each agent $i$ calculates a parameter called \textit{attractiveness} ($t_{i,a}$) defined as:
%
$t_{i,a}=(1-x_{i,a})h_a+x_{i,a}U_{i,a}$,
%
where $x_{i,a}$ is the \textit{attitude} of action $a$, which is initially selected randomly from $[0,1]$. Moreover, $h_a$ is the fraction of rounds in which action $a$ has won in a given history of the game. The action with the highest attractiveness is chosen by the agent in the next round of the play. As the game evolves, in each round of play, an agent adapts her attitude values such that if agent selects action $a$ and wins, $x_{i,a}$ will be increased by some constant $a+$ whereas if agent selected action $a$ and lost, $x_{i,a}$ will be decreased by some constant $a-$.
\item \textbf{Win-Stay Lose-Shift Strategy:} In \cite{REENTS2001253}, this learning method is presented as a simple behavioral model for the agents playing an MG. This is a stochastic strategy-based learning method. If an agent wins in the current round of the game, she selects the same action in the next round. In contrast, if the agent loses, she will choose the other action with some probability $p$. Authors analytically showed that for small enough $p$ values, the social welfare (i.e., volatility) of the system approaches the optimal value. More precisely, for the MG with $N$ odd players and $N/2$ cutoff value, $p$ is chosen such that $p=x/(N/2)$ where $x \ll N$.
\item \textbf{Roth-Erev Learning:} This learning method is applied in MG in \cite{Catteeuw2012}. Similar to the Q-learning, an agent determines a weight for each of her actions, denoted by $q_a$ and referred to as \textit{action weights}. However, unlike Q-learning, $q_a$ is defined as the sum of the initial action weight and the discounted sum of all past utility values received for playing action $a$ ($\lambda$ is referred to as the \textit{discount factor}). Agents use the following rule to update the actions' weight:
\begin{equation*} 
\label{automata}
q_a(t+1)= \begin{cases}
\lambda q_a(t)+U_{i,a}(t), \ a_{i,t}=a \\
\lambda q_a(t),\ \text{O.w.} 
\end{cases}
\end{equation*}
Given the values of $q_{a}$, the selection probability of action $a$ is defined as $p_a = \frac{q_a}{\sum_{a'}q_{a'}}$. 
\item \textbf{Learning Automata:} According to \cite{Catteeuw2012}, learning automata can be applied as an MG learning mechanism, by using the following rule to update the probability of playing every action $a$, denoted by $p_a$, after each round of play:
\begin{equation*}
\begin{aligned} 
&p_a(t+1)=\\ 
&\begin{cases}
p_a(t)+\gamma U_{i,a}(1-p_a(t))-\delta (1-U_{i,a})p_a(t), \ a_{i,t}=a \\
p_a(t)-\gamma U_{i,a} p_a(t) +\delta (1-U_{i,a})(\frac{1}{2}-p_a(t)),\ \text{O.w.}
\end{cases}
\end{aligned}
\end{equation*} 
Here $\gamma$ and $\delta$ are known as the \textit{reward rate} and \textit{penalty rate}, respectively.
\item \textbf{Random selection:} In a random selection scenario, agents simply select one of the two actions uniformly at random.
\end{itemize}
\subsection*{Numerical Results}
\label{simulation}
\begin{figure}[!htb]
\centering
\includegraphics[width=0.450 \textwidth]{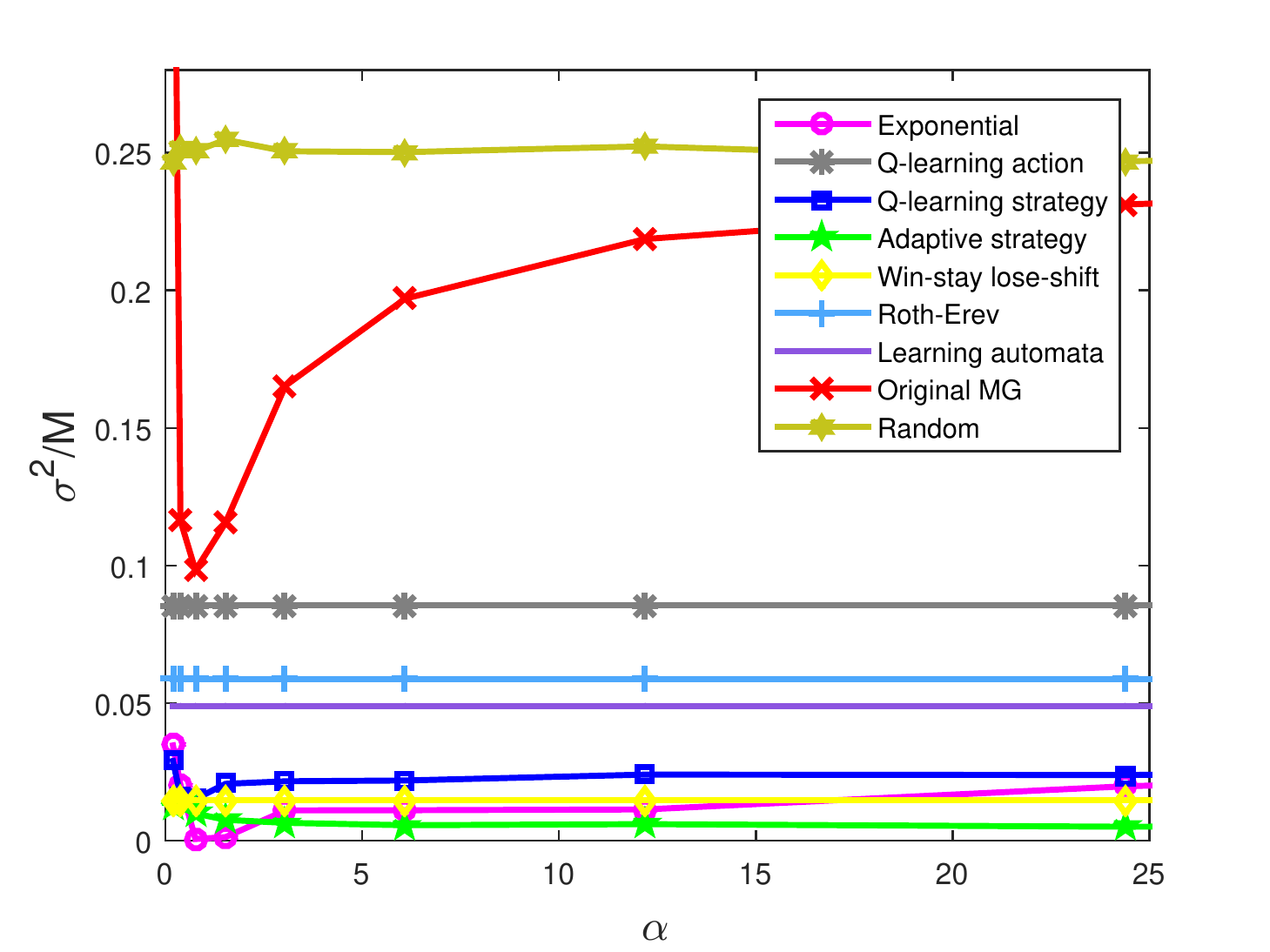}
\caption{Performance comparison in terms of (inverse) aggregate utility.}
\label{volatility}
\end{figure}

\begin{figure}[!htb]
\centering
\includegraphics[width=0.450 \textwidth]{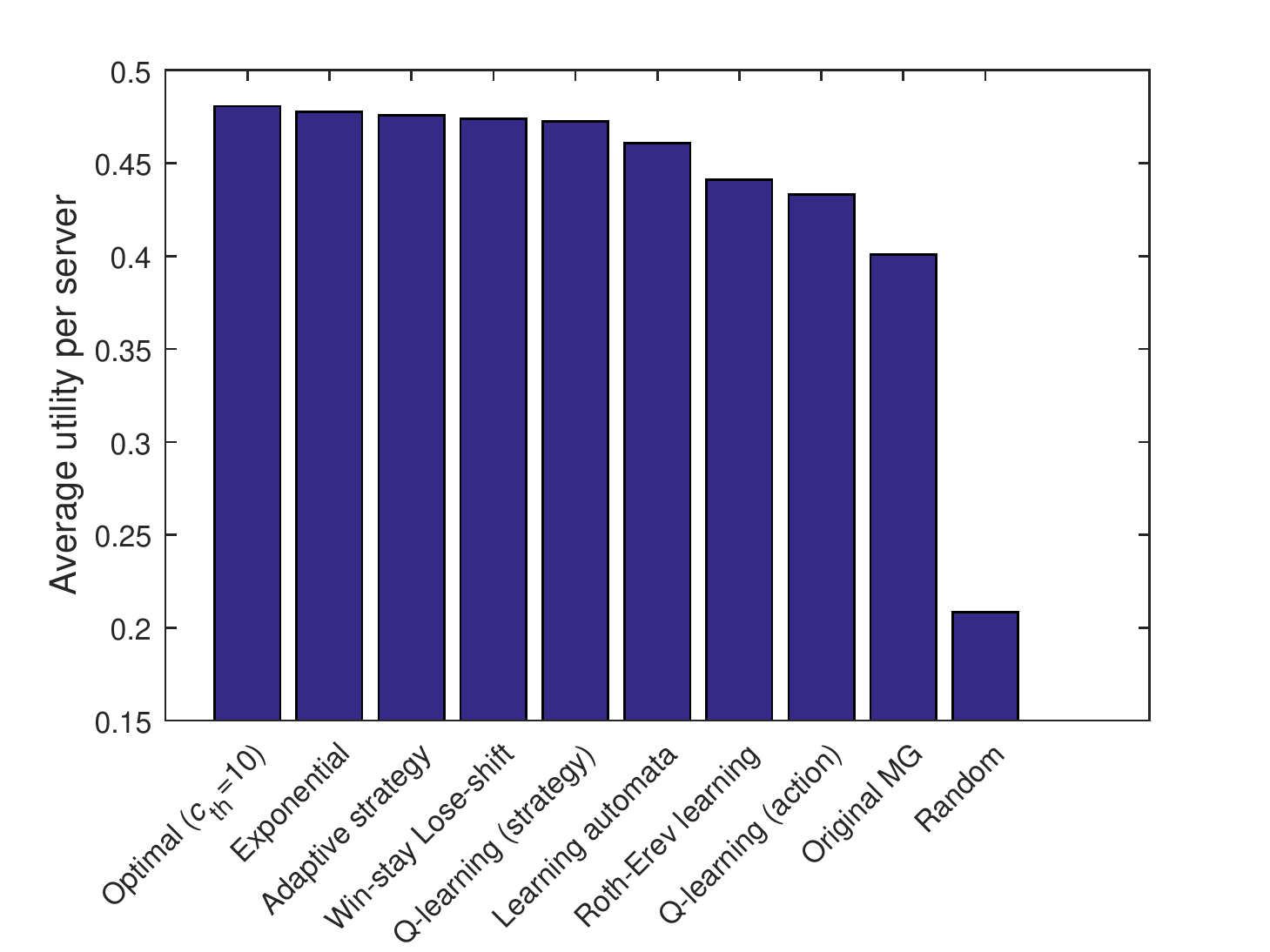}
\caption{Performance comparison in terms of average utility for each server.}
\label{utility}
\end{figure}

\begin{figure}[!htb]
\centering
\includegraphics[width=0.450 \textwidth]{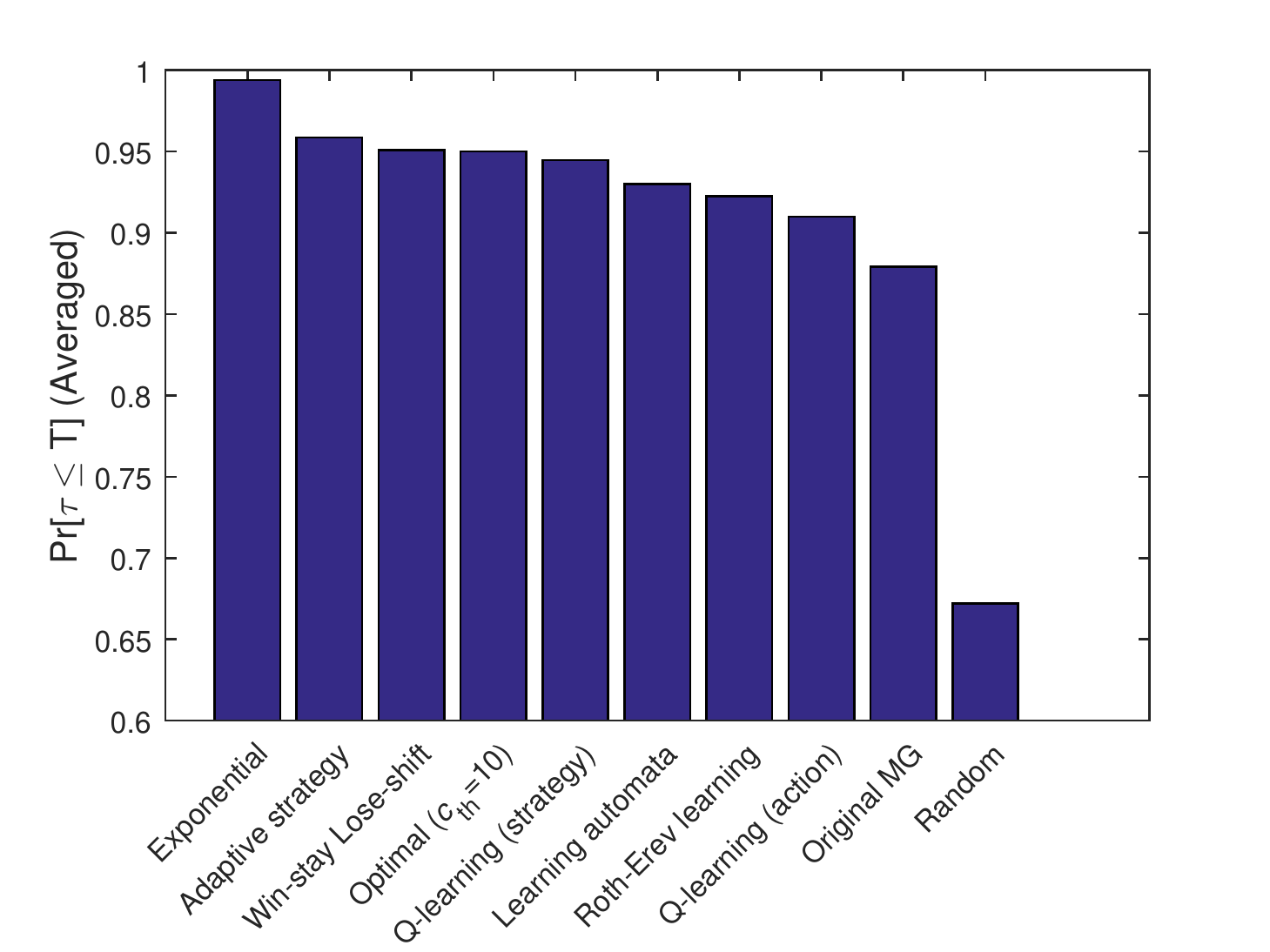}
\caption{Performance comparison in terms of users' experience.}
\label{probability}
\end{figure}

We choose $M=21$ and $c_{\textup{th}}=10$. Simulations are carried out for $32$ runs and in each run, the servers repeatedly execute the MG for $10000$ offloading periods. We compare all aforementioned learning methods based on the social- and individual welfare of servers as well as users' QoE measure. For different learning schemes, the parameters are selected as follows, using the best values as suggested in the literature:
\begin{itemize}
\item Exponential learning: $\gamma=100$.
\item Q-learning: $\gamma=0.1$, $\epsilon=0.01$.
\item Adaptive strategy: Initial attitude values $x_{i,0}=x_{i,1}=0.5$, and $a+=a-=0.5$, $\forall i \in \{1,...,M\}$. 
\item Win-stay lose-shift strategy: $p=0.005$.
\item Roth-Erev learning: $\lambda=0.2$.
\item Learning automata: $\gamma=0.2$ and $\delta=0.3$.
\item Seminal MG: $S=2$.
\end{itemize}
In Fig. \ref{volatility}, we show the variations in the volatility as a function of the parameter $\alpha=2^s/M$, with $s$ being the 
\textit{memory size}, i.e., the length of the historical data used by the agents for learning. It can be seen that exponential learning method achieves the best social welfare (inversely proportional to the volatility), with its lowest volatility approaching to $0$.  
In addition to examining the social welfare of the system, we also investigate the performance of each learning method in terms of individual welfare of the servers. In doing so, we illustrate the average utility per server during the entire the game in Fig. 
\ref{utility}. It can be concluded that using an appropriate learning method, a near-optimal average utility is achievable by the servers, despite not having any prior information. 
Finally, in Fig. \ref{probability}, we compare the performance of different learning methods in terms of users' experience. To this end, we simulate the probability that the total time required by each server to perform all tasks, denoted by $\tau$, exceeds the deadline $T$. Naturally, $\Pr[\tau \leq T]$ is then the probability that no user, even the last one in the queue, would experience a delay larger than 
$T$ to have its offloaded task done. 

The learning methods such as exponential learning, adaptive strategy, win-stay lose-shift strategy and Q-learning that exhibit better performance than the seminal inductive learning method help servers achieve better coordination and thus form larger minorities. This reduces wastage of computation server resources and hence improves the resource allocation efficiency. Therefore, these methods can be recommended as more efficient and sophisticated learning rules for the formulated MG-based server selection problem.
\subsection*{Extension of the Model}
\label{open_issues}

Since MEC networks typically consist of a variety of edge nodes such as small base stations, macro base stations, wireless access points, etc., the edge servers are not homogeneous in practice. Therefore, heterogeneities in their computational capability, power and storage should be taken into account when developing efficient resource allocation mechanisms. To model such scenarios, games that incorporate different types of players could be applied. In addition, to ensure fairness among the servers, analyses using various equilibrium notions need to be carried out. Moreover, mathematical tools such as queuing theory and Markov decision processes can be used to more accurately model the randomness in the offloading system such as random arrival of computation tasks and the users' status change. 
\section*{Conclusion}
\label{Sec:conclusion}
We have outlined the major challenges that arise in MEC, primarily focusing on computation offloading. We have investigated the state-of-the-art and studied the applicability of distributed solution approaches such as game theory and reinforcement learning for deriving efficient solutions for the identified challenges. Moreover, we have formulated the energy efficient edge server activation problem in a MEC offloading system using minority games and obtained some preliminary results by applying a number of reinforcement learning techniques. Extension of the model to consider several practical aspects of the efficient resource management problem for MEC servers has also been discussed. 
%
\bibliographystyle{IEEEtran}
\bibliography{References}
\end{document}